\documentclass[twocolumn,showpacs,preprintnumbers,longbibliography,prb]{revtex4-1}
\usepackage{graphicx,bm,amsmath,amssymb, ulem}
\usepackage{hyperref}
\hypersetup{
    colorlinks=true,
    linkcolor=blue,
    filecolor=magenta,      
    urlcolor=blue,
}

\usepackage{xcolor}
\usepackage{comment}

\def\gz{\ifmmode{Z\hskip -4.8pt Z}
    \else{\hbox{$Z\hskip -4.8pt Z$}}\fi}

\newcommand{\be}{\begin{equation}}
\newcommand{\ee}{\end{equation}}
\newcommand{\bea}{\begin{eqnarray}}
\newcommand{\eea}{\end{eqnarray}}

\begin{document}

\title{Magnon-assisted dynamics of a hole doped in a cuprate superconductor}
\author{I. J. Hamad}
\affiliation{Instituto de F\'{\i}sica Rosario (CONICET) and Facultad de Ciencias Exactas, Ingeniería y Agrimensura, Universidad Nacional de Rosario, 
Bv. 27 de Febrero 210 bis, 2000 Rosario, Argentina}
\author{L. O. Manuel}
\affiliation{Instituto de F\'{\i}sica Rosario (CONICET) and Facultad de Ciencias Exactas, Ingeniería y Agrimensura, Universidad Nacional de Rosario, 
Bv. 27 de Febrero 210 bis, 2000 Rosario, Argentina}
\author{A.~A.~Aligia}
\affiliation{Centro At\'{o}mico Bariloche and Instituto Balseiro, Comisi\'{o}n Nacional de Energ\'{\i}a At\'{o}mica, CONICET, 8400 Bariloche, Argentina}

\begin{abstract}
We calculate the quasiparticle dispersion and spectral weight of the quasiparticle that results when a hole is added to 
an antiferromagnetically ordered CuO$_2$ plane of a cuprate superconductor. We also calculate the magnon contribution to the quasiparticle spectral function. 
We start from a multiband model for the cuprates considered previously [Nat. Phys. \textbf{10}, 951 (2014)].
We map this model and the operator for creation of an O hole to an effective 
one-band generalized $t-J$ model, without free parameters.
The effective model is solved using the state of the art self-consistent Born approximation.
Our results reproduce all the main features of experiments. They also reproduce qualitatively the dispersion of the multiband model, 
giving better results for the intensity near wave vector $(\pi,\pi)$, in comparison with the experiments. 
In contrast to what was claimed in [Nat. Phys. \textbf{10}, 951 (2014)], we find that spin fluctuations play an essential role in the dynamics 
of the quasiparticle, and hence in both its weight and dispersion. 
\end{abstract}

\maketitle

\date{\today }

\section{Introduction}
\label{intro}
More than three decades after the discovery of high temperature superconductors, the issue of the appropriate 
microscopic minimal model that correctly describes the low-energy physics is still debated. 
There is however a consensus on the validity the three-band model $H_{3b}$  for describing the physics of the cuprates
at energies below $\sim 1$ eV, where the three bands come from two 
O 2p$_{\sigma }$ orbitals (those pointing in the direction of the nearest Cu sites) and one Cu 3d$_{x^2-y^2}$ 
orbital.\cite{eme,varma} 
At higher energies other orbitals should be considered.\cite{rai,raman,jiang}  
Other models used to describe the cuprates are the spin-fermion model $H_{sf}$,\cite{eme2} obtained from $H_{3b}$ 
after eliminating the Cu-O hopping by means of a canonical transformation (only the $d^9$ configuration of Cu is 
retained, represented by a spin $1/2$, which interacts with the fermions of both O bands),\cite{sf1,sf}
and the generalized $t-J$ model $H_{GtJ}$,\cite{zr,sys} which consists of holes moving in a background of Cu $1/2$ spins  
with antiferromagnetic exchange $J$, nearest-neighbor hopping $t$, and additional terms of smaller magnitude.

$H_{GtJ}$ is derived as a low-energy effective one for $H_{3b}$ or $H_{sf}$,\cite{sys,bel,fei} assuming that the 
low-energy part of the multiband models is dominated by Zhang-Rice singlets (ZRSs),\cite{zr} which 
in $H_{sf}$ consist of singlets formed between the spin of a cooper atom and the spin of the hole residing in a 
linear combination $L$ of four ligand oxygen orbitals around the cooper atom.\cite{zr,sys} 
In  $H_{3b}$, in which charge fluctuations are allowed, the ZRS also includes states with two holes in the Cu 
3d$_{x^2-y^2}$ orbital and in the O $L$ orbital.\cite{fei,comm} 
The proposal of Zhang and Rice has initiated a debate about the validity of a 
one-band model that continues at 
present.\cite{emerei,zha,ding,rc,cuge,lau,laucom,ebrahimnejad14,ebrahimnejad16,broo,Chainani17,Adolphs,tcuo}.
This issue is of central importance since models similar to the $t-J$ model were used to explain many properties of 
the cuprates,\cite{Greco09,Greco19} including superconductivity.\cite{manu,mal,Plakida}

An important probe for the models is the spectral function of a single-hole doped on the parent half-filled  
compounds, whose quasiparticle (QP) dispersion relation is directly measured in angle-resolved photoemission 
(ARPES) experiments.\cite{wells95,Moser} Experimental evidence shows that this doped hole resides mainly on the  O 2p$_{\sigma }$ 
orbitals.\cite{nuc,taki,oda} 
Naively one might expect that this fact is a serious problem
for $H_{GtJ}$, since O holes are absent in the model. However, mapping appropriately the corresponding operators, Cu and O 
photoemission spectra can be calculated with both $H_{sf}$,\cite{sf} and $H_{GtJ}$.\cite{eroles99} Nevertheless, while the 
experimental dispersion observed in Sr$_2$CuO$_2$Cl$_2$,\cite{wells95} has been fit using $H_{GtJ}$,
an unsatisfactory aspect is that the ``bare'' $t-J$ model with only nearest-neighbor hopping $t$ was unable to explain the 
observed dispersion, and {\it ad hoc} hopping to second and third nearest neighbors were included.\cite{naza,xiang,beli2,lema97}

In Ref. \onlinecite{ebrahimnejad14}, the QP dispersion $E_{QP}({\bf{k}})$ and its intensity $Z_{QP}({\bf{k}})$ for adding an O hole in an undoped CuO$_2$ 
plane were calculated, using $H_{sf}$ solved with an approximate variational method using realistic parameters. The dispersion obtained 
agrees with experiment. However, the reported intensity increases as $k$ moves from $(\frac{\pi}{2},\frac{\pi}{2})$ to $(\pi,\pi)$, in contrast 
to experiment. In the mentioned reference, it was also claimed that spin fluctuations play a minor role in the dynamics of the hole.

In this work we map the $H_{sf}$ used in Ref. \onlinecite{ebrahimnejad14} to an $H_{GtJ}$ without adjustable parameters, extending 
to $Z_{QP}$ the procedure we used before for T-CuO.\cite{tcuo} The resulting $H_{GtJ}$ is solved using the state of the art 
self-consistent Born approximation (SCBA). We obtain results in agreement with experiment for both $E_{QP}$ and $Z_{QP}$.
We also calculate the hole's spectral function, by taking into account multimagnon contributions within the SCBA.  
In this way we argue that the spin fluctuations play an essential role in the hole's dynamics.  
In particular the width of $E_{QP}$ is determined by the nearest-neighbor spin exchange $J$.

\section{Spin-fermion model and the one-band model derived from it}

\label{models}
\begin{figure}
\vspace*{0.cm}
\includegraphics[width=0.48\textwidth]{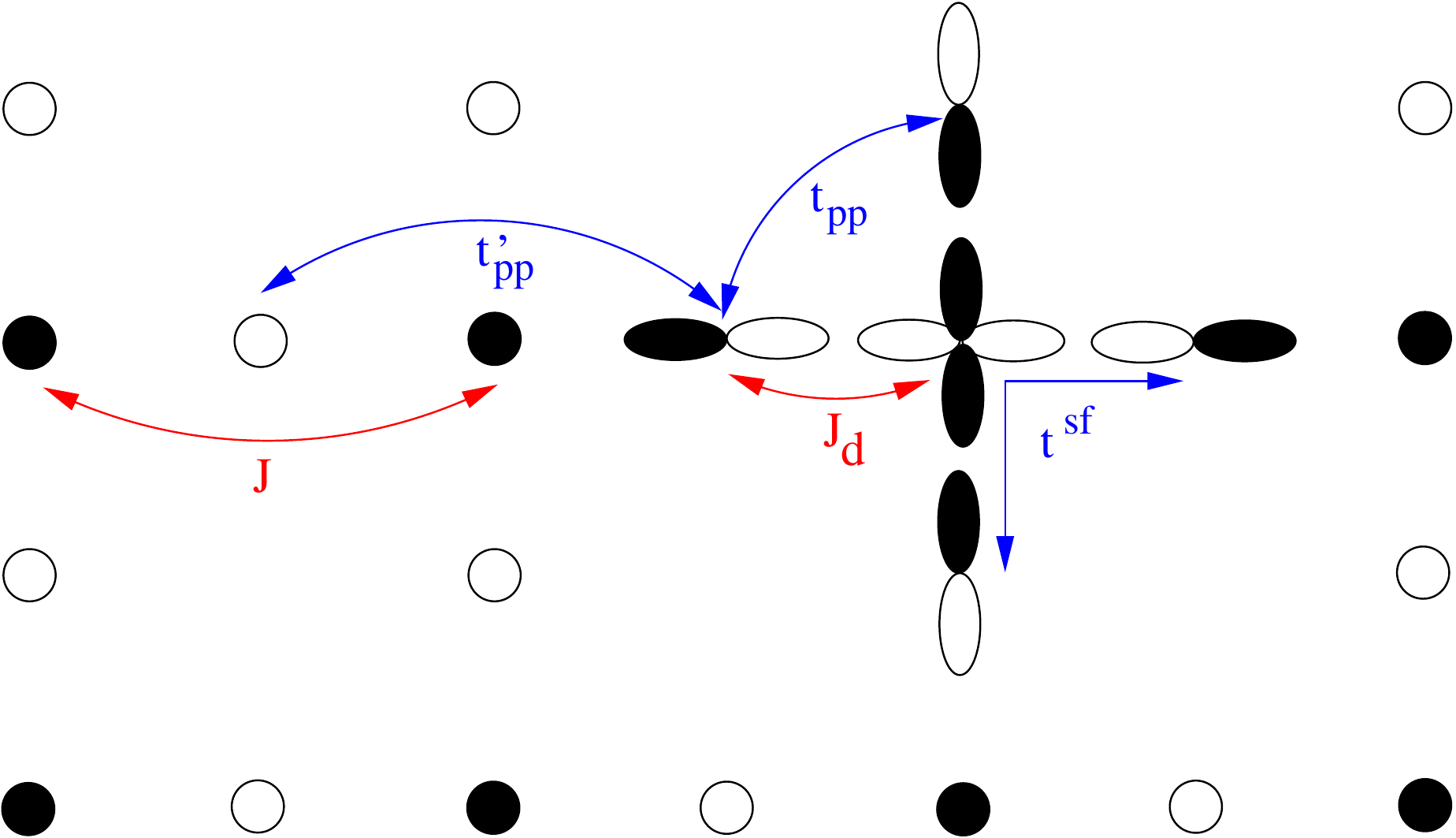}
\caption{Structure of the CuO$_2$ planes and sketch of the parameters of the 
three-band spin fermion model [Eq. (\ref{hsf})]. Filled (empty) circles represent Cu (O) sites.}
\label{red}
\end{figure}

We start from the spin-fermion model (Cu spins and O holes), obtained from $H_{3b}$ integrating out valence fluctuations at the 
Cu sites.\cite{eme2,sf1,sf,ebrahimnejad14} 
With the adequate choice of phases the Hamiltonian reads

\begin{eqnarray}
H_{sf} &=&\sum_{i\delta \delta ^{\prime }\sigma }p_{i+\delta ^{\prime}\sigma }^{\dagger }p_{i+\delta \sigma }
\left[ (t_{1}^{sf}+t_{2}^{sf})(\frac{1}{2}+2\mathbf{S}_{i}\cdot \mathbf{s}_{_{i+\delta }})-t_{2}^{sf}\right]\notag \\ 
&&-t_{pp}\sum_{j\gamma \sigma }p_{j+\gamma \sigma }^{\dagger }p_{j\sigma}+t_{pp}^{\prime }\sum_{j\delta \sigma }
\left( p_{i+\delta \sigma }^{\dagger}p_{i-\delta \sigma }+\mathrm{H.c.}\right)   \notag \\
&&-\sum_{i\delta} J_{d}\mathbf{S}_{i}\cdot \mathbf{s}_{_{i+\delta }}+
\frac{J}{2}\sum_{i\delta }\mathbf{S}_{i}\cdot \mathbf{S}_{_{i+2\delta }},
\label{hsf}
\end{eqnarray}
where $i$ $(j)$ labels the Cu (O) sites, $i+\delta $ ($j+\gamma )$ label the four O atoms nearest to Cu atom $i$ (O atom $j$), and 
$p_{j\sigma}^{\dagger }$ creates an O hole at the 2p$_{\sigma }$ orbital of site $j$ with spin $\sigma $. 
The spin at the Cu site $i$ (O orbital 2p$_{\sigma }$ at site $i+\delta $) is denoted as $\mathbf{S}_{i}$ ($\mathbf{s}_{_{i+\delta}}$). 
As in Ref. \onlinecite{ebrahimnejad14} we include hopping $t_{pp}^{\prime }$ between second-neighbor O orbitals with a Cu in between, 
and $J_{d}$ (which reduces part of the first term for $\delta ^{\prime }=\delta $)  not included in earlier studies. The model is represented in Fig. \ref{red}. 
In units of the Cu-Cu spin exchange $J=1$, the parameters chosen for the multiband model of Ref.~\onlinecite{ebrahimnejad14} 
are: $t_{1}^{sf}=2.98$, $t_{2}^{sf}=0$, $t_{pp}=4.13,$ $t_{pp}^{\prime }=2.40$, and $J_{d}=3.13$.

Projecting the Hamiltonian over the subspace of ZRS's, we have derived a one-band generalized $t-J$ model:  
\begin{equation}
H_{GtJ} =-\sum_{\kappa }t_{\kappa }\sum_{iv_{\kappa }\sigma }\left(
c_{i\sigma }^{\dagger }c_{i+v_{\kappa }\sigma }+\mathrm{H.c.}\right)+\frac{J}{2}\sum_{iv_{1}}\mathbf{S}_{i}\cdot \mathbf{S}_{_{i+v_{1}}},  \label{hgtj}
\end{equation}
where \ $c_{i\sigma }^{\dagger }$ creates a hole at the Cu site $i$  with spin $\sigma $, and $\kappa =1,2,3$, refer to first, second, and third 
nearest-neighbors $v_\kappa$  within the  sublattice of Cu atoms. Additional terms are small and do not affect the hole dynamics. 
The derivation of this one-band Hamiltonian and the calculation of its parameters follow the procedure detailed in the supplemental material of Ref. \onlinecite{tcuo}, here generalized to include 
the effect of second nearest-neighbor O hopping $t_{pp}^{\prime }$. The contribution of this term for a hopping $\tau _{R}$ between sites at a distance $R=(x,y)$ becomes:    
\begin{eqnarray}
\tau _{R} &=&\frac{2t_{pp}^{\prime }}{N}\sum_{k}\cos (k_{x}x)\cos (k_{y}y) 
\notag \\
&&\times \left( 1-\frac{\cos ^{4}(k_{x}b)+\cos ^{4}(k_{y}b)}{\cos
^{2}(k_{x}b)+\cos ^{2}(k_{y}b)}\right) ,  \label{tau}
\end{eqnarray}
where $b=a/2$ is half the lattice parameter and $N$ is the number of sites of the cluster. The contribution of the other terms of $H_{sf}$ to the different 
terms of $H_{GtJ}$ has been described in detail before.\cite{tcuo}. The resulting parameters 
of $H_{GtJ}$ are, taking $J=0.15$ eV as the unit of energy, 
$t_{1}=1.921$, $t_{2}=-0.371$, $t_{3}=0.592$. 

\section{Treatment of the one-band model}

\label{scba}

We calculate the QP spectral functions --from which the single hole's dispersion and weight are directly derived-- 
and the magnon contributions to the hole's wave function (WF) by means of the SCBA,~\cite{martinez91,lema97,lema98,Trumper04} 
a semianalytic method that compares very well  with exact diagonalization (ED) results on small clusters in different 
systems.~\cite{martinez91,lema97,Trumper04,Hamad08,Hamad12} 
It is one of the most reliable and checked methods up to date to calculate the hole Green's function,  and in particular its 
QP dispersion relation. However, some care is needed to map the QP weight between different models.~\cite{lema97} 
In order to do such calculation, we follow standard procedures.~\cite{martinez91} 
On one hand, the magnon dispersion relation is obtained treating the magnetic part of the Hamiltonian at the linear spin-wave level, 
since the system we study has long-range antiferromagnetic order, and it is well known that its magnetic excitations are 
semiclassical magnons.~\cite{coldea01} 
On the other hand, the electron creation and annihilation operators 
in the hopping terms are  mapped into holons of a slave-fermion representation (details in Ref. \onlinecite{tcuo}). 
Within SCBA, we arrive to an effective Hamiltonian: 
\begin{eqnarray}
H_{\text{eff}} &=&\sum_{\mathbf{k}}\epsilon _{\mathbf{k}}h_{\mathbf{k}}^{\dagger
}h_{\mathbf{k}}+\sum_{\mathbf{k}}\omega _{\mathbf{k}}\alpha _{\mathbf{k}%
}^{\dagger }\alpha_{\mathbf{k}}+ \nonumber \\
&&+\frac{1}{\sqrt{N}} \sum_{\mathbf{kq}%
}\left(M_{\mathbf{kq}}h_{\mathbf{k}}^{\dagger }h_{\mathbf{k}-\mathbf{q}}\alpha_{%
\mathbf{q}}+\mathrm{H.c.}\right), 
\end{eqnarray}
with
\begin{equation}
\epsilon _{\mathbf{k}} = 4t_{2}\cos (ak_{x})\cos (ak_{y})+ 2t_{3}\left[ \cos (2ak_{x})+\cos (2ak_{y})\right],
\nonumber
\end{equation}
\begin{equation}
\omega_{\bf{k}} = \sqrt{A_{\bf{k}}^{2}-4B_{\bf{k}}^{2}},  
\nonumber
\end{equation}
\begin{equation}
M_{\mathbf{kq}} =  2t_{1}\left[ u_{\mathbf{q}}\zeta (\mathbf{k-q})-v_{\mathbf{q}%
}\zeta (\mathbf{k})\right],   
\label{HeffSCBA}
\end{equation}

\normalsize

\noindent where $\epsilon _{\mathbf{k}}$ is the bare hole dispersion (with no coupling to magnons), 
$\omega_{\bf{k}}$ is the magnon dispersion relation, with 
$A_{\bf{k}} =2J$, $B_{\bf{k}}=\frac{J}{4}\sum_{\bf{R}}\cos ({\bf{R}}\cdot \bf{k})$, 
and $M_{\mathbf{kq}}$ is the vertex that couples the hole with magnon excitations. 
Here $\zeta (\mathbf{k}) =\cos (ak_{x})+\cos (ak_{y})$ being $a$ the distance between Cu atoms in the CuO$_{2}$ planes and where $u_{\mathbf{q}}$ and $v_{\mathbf{q}}$ are the usual Bogoliubov coefficients. 

The heart of the SCBA method lies in the self-consistent Dyson equation for the hole's self energy~\cite{kane89} 
$$\Sigma_{\bf k}(\omega )=\frac{1}{N}\sum_{\bf q}\mid M_{\bf kq}\mid^{2} G_{{\bf k}-{\bf q}}(\omega-\omega_{\bf q}),$$
being $ G_{\bf k}(\omega)=(\omega -\epsilon_{\bf k} -\Sigma_{\bf k}(\omega))^{-1}$ the hole Green's function. 
From the self-energy the QP energy can be computed, by means of the self-consistent equation $E_{QP}({\bf k})=\Sigma_{\bf k}(E_{QP}({\bf k}))$, and also the holon spectral weight, defined as~\cite{kane89}

\begin{equation}
 Z_h({\bf k}) = \left.\left(1-\frac{\partial {\rm Re}\;\Sigma_{\bf k} (\omega)}{\partial \omega}\right)^{-1}\right|_{E_{QP}({\bf k}}).
 \label{zhk}
\end{equation}

Although Eq. (\ref{zhk}) in principle allows the calculation of the spectral weight directly, in practice within the SCBA it is impossible to apply it due to the strong irregularities in the derivative of $ {\rm Re} \Sigma_{\bf k}$. Instead, the spectral weight is calculated by integrating the QP peak in the spectral function.

\subsection{QP spectral function and magnon coefficients of the QP wave function}

For the calculation of the magnon contributions to the QP's WF we follow the steps taken in 
Refs.~\onlinecite{reiter94,ramsak98,Trumper2004}. The QP WF with momentum ${\bf{k}}$ can be expressed as a sum of terms, each of which involves the contribution of a growing number of magnons.  
Hence, within the SCBA, the QP WF results by taking the $n \to \infty$ limit of: 
$$
|\Phi^{n}_{\bf k}\rangle =  Z_h({\bf k})\left[ h^{\dagger}_{\bf k}
+ \frac{1}{\sqrt{N}}\sum_{{\bf q}_{1}}g_{{\bf k},{\bf q}_{1}}
h^{\dagger}_{{\bf k}-{\bf q}_{1}} \alpha^{\dagger}_{{\bf q}_1}
+... \right.$$
$$
\left. +\frac{1}{\sqrt{N^n}} \sum_{{\bf q}_{1},.....,{\bf q}_{n}}
g_{{\bf k},{\bf q}_{1}} g_{{\bf k}_1,{\bf q}_{2}}....
g_{{\bf k}_{n-1},{\bf q}_{n}}
h^{\dagger}_{{\bf k}_n}
\alpha^{\dagger}_{{\bf q}_1}...\alpha^{\dagger}_{{\bf q}_n} \right] |AF\rangle,
$$
where ${\bf k}_i={\bf k}-{\bf q}_1-...-{\bf q}_i$, $|AF\rangle$ is the undoped antiferromagnetic ground state, and
\begin{equation}
g_{{\bf k}_{n},{\bf q}_{n+1}}= M_{{\bf k}_n,{\bf q}_{n+1}} G_{{\bf k}_{n+1}} (E_{QP}({\bf k})-\omega_{{\bf q}_1}-....-\omega_{{\bf q}_{n+1}}).
\label{greenk}
\end{equation}
It can be seen that each contributing term to the QP WF involves a growing number of magnons, starting from the first 
zero magnon term whose relative weight is given by the holon spectral weight $Z_h({\bf k})$. 

The QP WF satisfies the normalization condition 
\begin{equation}
 S_{\bf k} = \lim_{n \to\infty} \langle \Phi^{n}_{\bf k}|\Phi^{n}_{\bf k}\rangle=  \sum^{\infty}_{m=0} A^{(m)}_{\bf k}=1.
 \label{normalization}
\end{equation}
Each coefficient $A^{(m)}_{\bf k}$ is the $m$-magnon contribution to the QP WF and is defined as
\begin{equation}
A^{(m)}_{\bf k}=\frac{z_{\bf k}}{N^{m}}
\sum_{{\bf q}_{1},.....,{\bf q}_{n}}
g^2_{{\bf k},{\bf q}_{1}} g^2_{{\bf k}_1,{\bf q}_{2}}
......g^2_{{\bf k}_{m-1},{\bf q}_{m}},
\label{coeficients}
\end{equation}
while for the particular case
$m=0$, $A^{(0)}_{\bf k} \equiv  Z_h({\bf k})$. 
In this way, within the SCBA the relative weight of each $n$-magnon term for the spin polaron can be evaluated for a specific moment 
of the Brillouin zone. 

In order to estimate the effective number of magnons necessary to have a 
reliable QP WF we can find the minimum $n$ such that 
$ S^{(n)}_{\bf k}= \langle \Phi^{n}_{\bf k}|\Phi^{n}_{\bf k}\rangle=
 \sum^{n}_{m=0} A^{(m)}_{\bf k} \simeq 1,$ 
 within certain precision. 

\section{Results}

In this Section we present the SCBA calculations for $H_{G\; t-J},$ using the previously estimated parameters 
and the experimental value $J \equiv J_0=0.15$ eV.

\subsection{Quasiparticle dispersion relation}
\begin{figure}
\vspace*{0.cm}
\includegraphics*[width=0.48\textwidth]{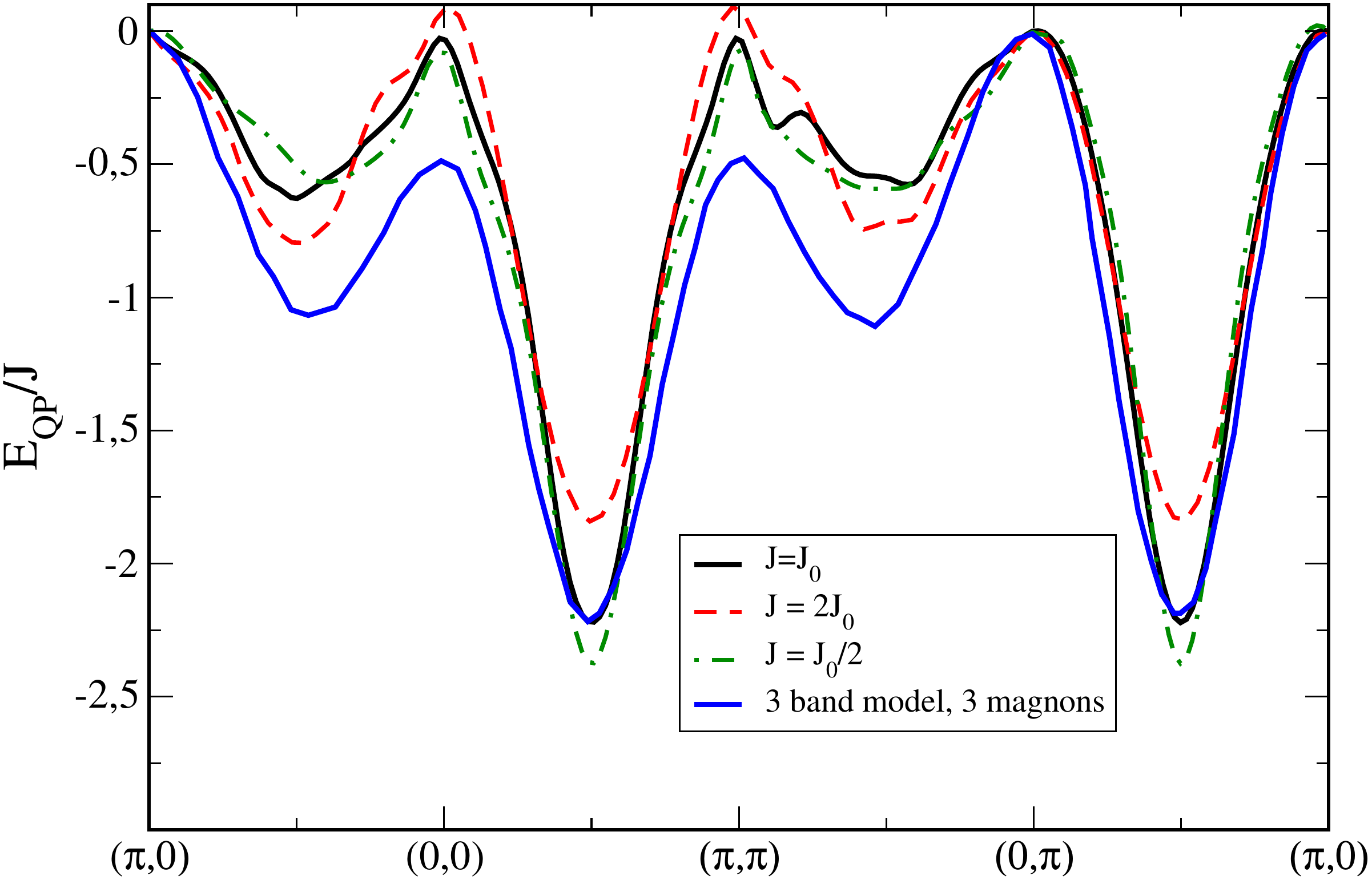}
\caption{QP hole dispersion relation. The solid black (blue) curve corresponds to the QP dispersion relation for the 
one-band generalized $t-J$ model (three-band model) calculated with SCBA (variational approach). The broken red (green) curve 
corresponds to the SCBA QP dispersion for a one-band model with an exchange interaction twice (half) the value of the 
experimental one.}
\label{reldisp}
\end{figure}

Fig. \ref{reldisp} shows the SCBA QP dispersion relation corresponding to our 
one-band generalized $t-J$ model along with the QP dispersion relation of the three-band model, obtained variationally (Ref. \onlinecite{ebrahimnejad16}). 
We recall that in our model there are no free parameters. All of them are rigorously obtained from the three-band model ~\cite{ebrahimnejad14} and experiments. The agreement of the one-band model and the multiband model dispersions is 
very good near the QP ground state momentum $\left(\frac{\pi}{2},\frac{\pi}{2}\right)$ and all along the diagonal and antidiagonal lines. In the rest of the chosen path, the agreement is semiquantitative. Compared with the ARPES measurements,~\cite{wells95,damascelli03} our results seem to better capture the quasi-degeneracy between the $(\pi,\pi)$ and $(\pi,0)$ points, with the energy at $(\pi,0)$ a little higher than at $(\pi,\pi)$. It should be stressed that, for simplicity, we are taking a hole picture, so the dispersion relation should be reversed in order to be 
compared with ARPES. From the dispersion relation only, it is not possible to conclusively discern whether the SCBA solution of the generalized $t-J$ model or the variational solution of the three-band model predictions agree better with ARPES. 

To analyze the role of the spin fluctuations for the hole motion within our theory, we also plot in Fig.~\ref{reldisp} the SCBA QP dispersion relation for the same $H_{G\;t-J}$ parameters but half and double exchange interaction $J$ values. The first point to be noticed is that as a first approximation, the bandwidth is directly proportional to $J$. When $J=2J_0$, that is, the spin fluctuations are enhanced in comparison with the hole kinetic energy, the relative dispersion bandwidth (in units of the corresponding $J$) is decreased, and now the energy of the ${\bf{k}}=(0,0)$ and $(\pi,\pi)$ points is slightly higher than that at the $(\pi,0)$ 
point, in contrast with ARPES. On the other hand, the dispersion for $J=J_0/2,$ i.e. when spin fluctuations are lowered, has 
the same structure as for $J=J_0$ but its relative bandwidth is larger than that of the three-band model. Therefore, it is evident that 
the spin fluctuations have a noticeable impact on the global dispersion form and its bandwidth. In particular, the increase of the 
exchange interaction gives rise to more localized QP states.

\subsection{Quasiparticle spectral weight}

\begin{figure}[ht]
\vspace*{0.cm}
\includegraphics*[width=0.48\textwidth]{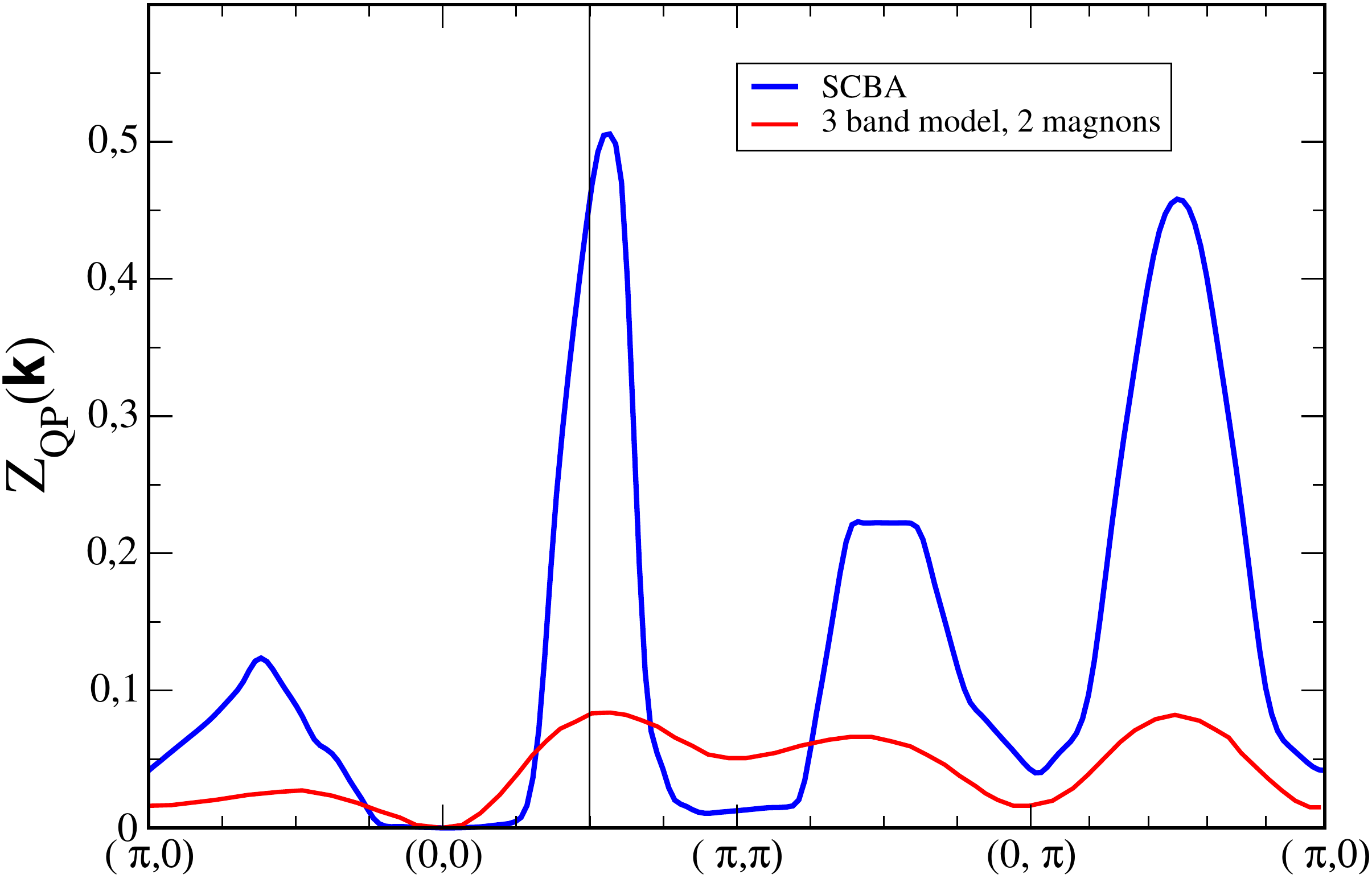}
\caption{QP spectral weight: the blue curve corresponds to the O contribution to the photoemission intensity calculated 
with SCBA, while the red curve corresponds to the QP weight function of the three-band model calculated variationally 
(taken from Ref.~\onlinecite{ebrahimnejad14}).}
\label{weight}
\end{figure}

In Fig.~\ref{weight} we show the QP spectral weight for the one-band and the three-band models along the same Brillouin zone path as in Fig.~\ref{reldisp}.  
Care must be taken in order to calculate the QP spectral weight within the one-band model, since almost all the contribution to the photoemission spectra comes from the addition of an O hole. However in the one-band model the O degrees of freedom have been integrated. 
In order to compute the O contribution to the ARPES QP intensity $Z_{QP}({\bf{k}})$ within the SCBA, we follow the procedures of Ref.~\onlinecite{eroles99}: we first calculate the {\it holon} spectral weight $Z_h({\bf k})$ , we then calculate the spectral weight for emitting a {\it physical} electron (see Ref. \onlinecite{lema97}), and finally from this we calculate the O intensity by means of a simple analytical relation between both, detailed in Ref.  ~\onlinecite{eroles99}.  In general, the calculated O intensity is higher than that of the variational calculation of the three-band model. Despite so, it can be seen that along the diagonal  $(0,0) \rightarrow (\pi,\pi)$, the intensity is large near the ground state $(\pi/2,\pi/2)$ momentum (note that it's not symmetric around $(\pi/2,\pi/2)$), but it decreases abruptly when approaching both $(0,0)$  and $(\pi,\pi)$. Nevertheless, those momenta do not show degeneracy in the intensity, as it happens for the holon weight within the SCBA~\cite{martinez91}. 

The general trend of the intensity calculated with the generalized $t-J$ model by means of the SCBA coincides with experiments,~\cite{wells95,damascelli03} in contrast with the results of the variational three-band model calculations~\cite{ebrahimnejad14}. In particular, the experiments show an almost vanishing QP photoemission weight close to $(0,0)$ and $(\pi,\pi)$  (see Fig. 1 of Ref.~\onlinecite{wells95}), that is correctly captured by our results, while in the three-band model calculation the $(\pi,\pi)$ point has an appreciable QP weight. In Ref.~\onlinecite{ebrahimnejad14} it was shown that appealing to a five-band model a partial decrease of the QP weight is obtained at $(\pi,\pi)$, while our more sophisticated SCBA calculation already captures this spectral feature in the one-band generalized $t-J$ model. Hence, we believe that the one-band model provides a quantitatively correct description of the photoemission spectra for the undoped cuprates.

\subsection{Magnon contributions to the QP wave function}
\begin{figure}
\vspace*{0.cm}
\includegraphics*[width=0.48\textwidth]{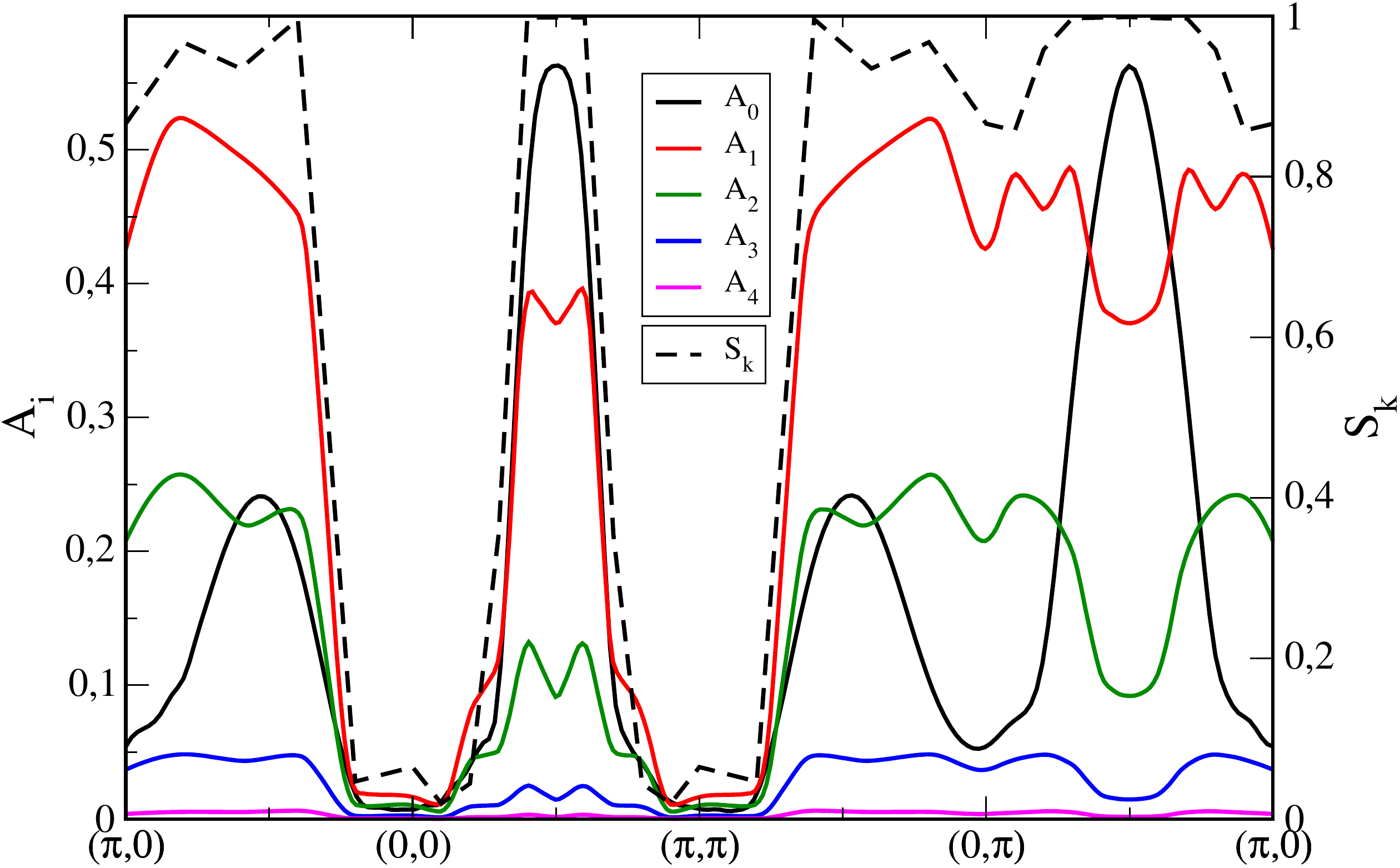}
\caption{Solid curves: Magnon coefficients $A^{(m)}_{\bf k}$ of the SCBA QP wave function for $N=400$. Broken curve: sum of the first 4 magnon coefficients.}
\label{magnones}
\end{figure}

In Fig.~\ref{magnones} we show the magnon coefficients $A^{(m)}_{\bf k}$ for $m=1, 2, 3$ and 4, along the same path in the 
Brillouin zone as in Fig.~\ref{reldisp}. The shown data were obtained for a cluster of $N = 400$ sites, using 25000 frequencies. 
We have checked that the results are essentially the same as for $N=1600$ sites, which is an indication that the $N=400$ cluster 
is a very good approximation for the thermodynamic limit.  We have chosen this cluster size because, for 1600 sites, the 
calculation of the fourth coefficient $A^{(4)}_{\bf k}$  is computationally expensive. 
For comparison, we put in Table~\ref{tablamagnones} the $A^{(m)}$ $m=1,2,3$ coefficients for the 1600 cluster, and for selected momenta 
along the diagonal $(0,0) - (\pi,\pi)$. 
It is worth to mention that for a correct computation of all the magnon coefficients is essential to get a very 
precise QP dispersion relation and its spectral weight, as can be seen from Eqs.~(\ref{greenk}) and (\ref{coeficients}). 
For this purpose, it is necessary to use a very large number of frequencies.
 
What can be clearly seen in Fig.~\ref{magnones} and Table~\ref{tablamagnones} is that the one- and the two-magnon coefficients can be, 
for many momenta, greater or of the same order of magnitude that the zero magnon coefficient $A^{(0)}_{\bf k}$, which we recall is the holon spectral 
weight $Z_h({\bf k})$. 
The three-magnon coefficient $A^{(3)}$ is small for all momenta but is by no means negligible. 
On the other hand, $A^{(4)}_{\bf k}$ is always very small, even compared to $A^{(3)}_{\bf k}$. 
From the magnon coefficients it can be concluded that spin fluctuations corresponding to several magnons are 
essential to build up the QP wave function. Since our one-band generalized $t-J$ model is rigorously derived from a 
multiband model and, as we have shown above, it reproduces the main features of the experimental QP dispersion relation and 
photoemission intensity, it can be stated that the spin polaron~\cite{martinez91} is the appropriate physical picture of the QP in 
cuprates.  

Fig.~\ref{magnones} also displays the partial sum of the norm $S^{(4)}_{\bf k}$, that is the sum of the first four magnon coefficients. 
It is evident that for those momenta where the holon QP weight $Z_h({\bf k})$ is not so small ($Z_h \gtrsim 0.05$), 
the normalization rule Eq. (\ref{normalization}) is reasonably satisfied with only up to a very few magnon coefficients. If the sum does not reach 
the value 1, it is very close, and hence it can be argued that with the inclusion of a few more magnon coefficients, the condition 
would be fulfilled. In this case, the QP can be thought as the bare hole moving around, exciting only up to three or four magnons. 
On the other hand, it is also clear that close to ${\bf{k}}=(0,0)$  and $(\pi,\pi)$, where the holon QP spectral weight is {\sout much } smaller than 0.05, 
the normalization condition is far from being satisfied. Since the four-magnon coefficients $A^{(4)}_{\bf k}$ are much smaller than the three-magnon ones $A^{(3)}_{\bf k}$, 
it is plausible to assume that the following coefficients would be smaller, and so there must be a ``magnon proliferation'', that is, 
the QP would be composed of a great number of magnons, and the sum rule can only be reasonably satisfied with a huge number of magnon 
coefficients, corresponding to very slowly convergent series.

\begin{table}[h]
\begin{center}
 \begin{tabular}{|c|c|c|c|c|c|c|}
\hline
{\bf{$k_x/\pi$}} & {\bf{$k_y/\pi$}} & $A^{(0)}_{\bf k}$ & $A^{(1)}_{\bf k}$ & $A^{(2)}_{\bf k}$ & $A^{(3)}_{\bf k}$ & $S_{\bf k}$\\
\hline
0.0 & 0.0 & 0.0048 & 0.0055 & 0.0032 & 0.00073 & 0.014 \\ 
0.1 & 0.1 & 0.0059 & 0.0080 & 0.0044 & 0.00095 & 0.019 \\ 
0.2 & 0.2 & 0.012 & 0.021 & 0.011 & 0.0023 & 0.047 \\ 
0.3 & 0.3 & 0.056 & 0.086 & 0.041 & 0.0087 & 0.19 \\ 
0.4 & 0.4 & 0.42 & 0.40 & 0.15 & 0.029 & 0.99 \\ 
0.5 & 0.5 & 0.55 & 0.38 & 0.10 & 0.017 & 1.00 \\ 
\hline
\end{tabular}
\caption{Magnon coefficients $A^{(m)}_{\bf k}$ for $m=1,2,3$ calculated for a $N=1600$ cluster, for selected momenta 
along the diagonal of the Brillouin zone. By symmetry, $A^{(m)}_{(\frac{\pi}{2}+k,\frac{\pi}{2}+k)}=A^{(m)}_{(\frac{\pi}{2}-k,\frac{\pi}{2}-k)}$.}
\label{tablamagnones}
\end{center}
\end{table}

In the pure $t-J$ model ($t_2 = t_3 = 0$), for a $J/t$ ratio as our $J/t_1$, Ramsak and Horsch~\cite{ramsak98} have shown that 
the QP is also composed of several magnons, and that for some momenta the one-magnon coefficient is larger than the zero-magnon one, and 
even that the two- and three-magnon terms are important to fulfill the normalization condition. This behaviour is analogous to the one we have found in
this work. However it is known that for the pure $t-J$ model the hole can only propagate by emitting and absorbing spin fluctuations~\cite{kane89}. 
Besides, it does not reproduce the experimentally measured dispersion~\cite{wells95}. With our generalized model, with second- and third-nearest neighbor 
hoppings, we were able to reproduce the experiments. It is usually argued~\cite{ebrahimnejad14} that, since $t_2$ or $t_3$ allow free hopping processes, 
in which the hole can move along a magnetic sublattice without disturbing the N{\'e}el order, the correct dispersion obtained by including 
further hoppings in the model implies that spin fluctuations do not play an important role in the QP formation. Our results indicate that this is 
not the case, and that for the generalized $H_{G t-J}$ the multimagnon processes are equally important in the formation of the QP as in the 
pure $t-J$ model. In previous works~\cite{Hamad08,Hamad12} we have already shown that even when there is a ``free hopping'' channel that allows
the hole to move without generating spin fluctuations of the magnetic background, the hole motion is promoted by emitting magnons, since this is
all in all energetically favourable.

\section{Conclusions}
Recent variational calculations~\cite{ebrahimnejad14,ebrahimnejad16,Adolphs} have suggested that one-band models cannot 
give a correct description of cuprates superconductors, based on the argument that these models, without {\it ad-hoc} terms, 
fail to describe even the ARPES photoemission spectra for a hole doped into an antiferromagnetically ordered CuO$_2$ layer. 
Also, these works have questioned a long-held belief about the spin polaron nature of a single hole doped in undoped 
cuprates.~\cite{kane89} To elucidate these claims, in this work we have performed a rigorous derivation of a one-band Zhang-Rice singlet based generalized $t-J$ model for cuprate superconductors, with no free parameters, starting from a 
three-band model. 
Its hopping terms, appreciable up to third nearest neighbors, are obtained from the three-band model 
parameters,~\cite{ebrahimnejad14} while the exchange interaction $J$ between copper sites is taken from experimental measurements. 

With the well established SCBA, we have computed the QP dispersion relation and the oxygen contribution to the photoemission intensity, 
obtaining a satisfactory agreement with ARPES experiments,~\cite{wells95,damascelli03} improving the above mentioned variational 
three-band model calculations~\cite{ebrahimnejad14}. Particularly, we have reproduced the experimental abrupt drop of the QP spectral weight 
going away from $(\frac{\pi}{2},\frac{\pi}{2})$ to $(\pi,\pi)$, that, within the variational calculation can only be partially obtained 
appealing to a more complicated five-band model. 

In addition, we have analyzed the structure of the SCBA QP wave function computing its magnon coefficients, and we have 
found that the spin fluctuations play an essential role in the building up of the QP. This happens even 
for our generalized $t-J$ model where second and third NN hoppings {\sout would} allow the hole motion without emitting magnon excitations of the 
antiferromagnetic background. 

From our results we can conclude that rigorously derived one-band models are appropriate for the description of 
(at least slightly doped) cuprate superconductors, while the physical nature of a single hole doped in CuO$_2$ layer 
corresponds to a spin polaron quasiparticle with spin fluctuations as its main ingredient.

\section{Acknowledgements.}

I. J. H and L. O. M. are partially supported by PIP 0364 of CONICET, Argentina. A. A. A. is sponsored by by PIP 112-201501-00506 of
CONICET, and PICT 2017-2726 and PICT 2018-01546 of the ANPCyT, Argentina.

\end{document}